\documentclass[notitlepage]{revtex4-1}

\usepackage{amsmath}
\usepackage{amsfonts}
\usepackage{amsthm}
\usepackage{amssymb}
\usepackage{amscd}
\usepackage{graphicx}
\usepackage{epsfig}
\usepackage{hyperref}

	\newcommand{\ncd}{\newcommand}
	\ncd{\mrm}    {\mathrm}
	\ncd{\beq}{\begin{equation}}
	\ncd{\eeq}{\end{equation}}
	\ncd{\nn}{\nonumber}

	\def\d{{\rm d}}

	\def\n{{\rm n}}
	\def\s{{\rm s}}

\begin{document}
\title{Variational thermodynamics of relativistic thin disks}

\author{Antonio C. Guti\'errez-Pi\~{n}eres}
\email[e-mail:]{acgutierrez@correo.nucleares.unam.mx}
\affiliation{Instituto de Ciencias Nucleares, Universidad Nacional Aut\'onoma de M\'exico,
 \\AP 70543,  M\'exico, DF 04510, M\'exico}
\affiliation{Facultad de Ciencias B\'asicas,\\
Universidad Tecnol\'ogica de Bol\'ivar, Cartagena 13001, Colombia}
		
\author{Cesar S Lopez-Monsalvo}
\email{cesar.slm@correo.nucleares.unam.mx}
\affiliation{Instituto de Ciencias Nucleares, Universidad Nacional Aut\'onoma de M\'exico,
 \\AP 70543,  M\'exico, DF 04510, M\'exico}

\author{Hernando Quevedo}
\email{quevedo@nucleares.unam.mx}
\affiliation{Instituto de Ciencias Nucleares, Universidad Nacional Aut\'onoma de M\'exico,
 \\AP 70543,  M\'exico, DF 04510, M\'exico}

\begin{abstract} 
We present a  relativistic model describing a  thin disk system composed  of  two  fluids. The system is  surrounded by a halo in the  presence  of a  non-trivial  electromagnetic field. We show that the model is compatible with the variational multi-fluid thermodynamics formalism, allowing us to determine all the thermodynamic variables associated with the matter content of the disk. The asymptotic behaviour of these quantities indicates that the single fluid interpretation should be abandoned in favour of a two-fluid model.  
\end{abstract}


\maketitle

The problem of finding exact solutions for the Einstein Field Equations  which are consistently applicable in the context of astrophysics remains a topical problem \cite{Bicak2002}. Most systems of astrophysical relevance are studied through various assumptions of symmetry. Of special interest are those which are approximately axially symmetric,  such as rotating compact objects and  accretion disks and galaxies in thermodynamic equilibrium. In the  case of  compact objects, the exterior field  is  usually assumed to be  described by  the  Kerr solution and the  interior  counterpart could be  described, for  instance, by  the  thin rotating dust disk constructed  by  Neugebauer and Meinel \cite{neug}. On the contrary, accretion disks and  galaxies  can  also  
 be approximated by a thin disk.  The conventional treatment of galaxies modelled as a thin disk has been largely studied using Newtonian dynamics. However, there are only a handful of physical solutions which are mathematically simple and fully relativistic. Moreover, most efforts in the understanding of the physical properties of such objects rely on the input provided through an equation of state.  In this work we present a new exact solution for a relativistic thin disk surrounded by an electromagnetic halo. This solution has a number of interesting features. Firstly, it is notoriously simple in its mathematical form, making it useful for testing various matter models in a straightforward manner. Secondly, it generalises the commonly used  pressure free (dust) models to a perfect fluid with non-vanishing pressure, allowing a more detailed physical description \cite{GG-PO,GG-PV}. Thirdly, we make a novel analysis by considering a multi-fluid system to describe the thermodynamics associated with the matter content of the solution. In this manner, we use a new thermodynamic criterion to exclude some dynamically unconstrained models in favour of those which are thermodynamically sound. Moreover, the  multi-fluid formalism is suitable to extend the analysis beyond equilibrium thermodynamics, providing us with a tool to test new directions in the context of relativistic astrophysics.

To obtain the solution  we solved the distributional Einstein-Maxwell field equations assuming axial symmetry and that the derivatives of the metric and electromagnetic potential across the disk space-like hyper-surface are discontinuous. Here, the energy-momentum tensor is taken to be the sum of two distributional components, the purely electromagnetic (trace-free) part and a `material' (trace) part.  
Accordingly, the Einstein-Maxwell
equations, in geometrized units such that $c = 8\pi G = \mu _{0} = \epsilon _{0} =  1$,
are equivalent to the system of equations
     \begin{align}
      G_{ab}^{\pm} =R^\pm_{ab} & - \frac{1}{2} g_{ab} R^\pm = E^\pm_{ab} + M^\pm_{ab},
      \label{eq:einspm}\\
      H_{ab} & - \frac{1}{2} g_{ab} H = Q_{ab}, \label{eq:einsdis}\\
      &\hat F^{ab}_{ \pm  \ \ , b}   =   \hat J^{a}_\pm ,\label{eq:maxext}\\
      &\  [\hat F^{ab}]n_{_{b}} =  \hat I^{a},\label{eq:emcasj}
      \end{align}
where $H = g^{ab} H_{ab}$, $\hat a = \sqrt{-g}a$ and $g$ the determinant of the metric tensor. Here, the square brackets in  expressions such as $[\hat F^{ab}]$ denote the jump of $\hat F^{ab}$ across of the surface $z=0$ and $n_{_{b}}$ denotes a unitary vector in the  direction normal  to it. 

 We solve the Einstein-Maxwell system  (c.f. \ref{eq:einspm}  -  \ref{eq:emcasj}) in a conformastatic space-time background through the introduction of an auxiliary harmonic function that determines the functional dependence of the metric components and the electromagnetic potential (c.f. Secion II in \cite{G-PGQ2013}). Inspired by inverse method techniques, let us assume that the solution has the general form
	\beq
	\label{eq:met0}
	\d s^2 = -e^{2 \phi} \d t^2 + e^{-2\beta \phi} \left\{r^2 \d \varphi^2 + \d r^2 + \d z^2 \right\},
	\eeq
where the metric function $\phi$ depends only on $r$ and $z$,  $\beta$ is taken to be a constant and the electromagnetic potential, $A_{\alpha} = (A_0, A, 0, 0)$,  is time-independent.  In order to  analyse the physical characteristics of the  system, it is convenient to work out all the relevant quantities in terms of the orthonormal tetrad of the ``locally static  observers'' (LSO) \cite{KBL}, i.e. observers at rest with respect to infinity.  The non-zero  components   of  the energy  momentum tensor of  the  halo
In  terms  of  the metric functions  (\ref{eq:met0})   are given in the  Appendix. Whereas the  current density  on the  halo as observed by a LSO  is  given 
by \begin{align}
 J_{(0)}^{\pm} 	&= e^{\beta\phi}\nabla\cdot \{ e^{-(1+\beta)\phi}\nabla A_0\}   \label{eq:current0},\\
 J_{(1)}^{\pm} 	&= 	-r e^{-\phi}\nabla\cdot \{ r^{-2}e^{(1+\beta)\phi}\nabla A\}  \label{eq:current1}.
	\end{align}
The  non-zero  components   of  the surface  energy--momentum  tensor  of  disk and  the  surface  electric  current  density  are given by 
\begin{align}
	S_{(0)(0)} 	&= 	4\beta e^{\beta\phi}\phi_{,z},\label{eq:stet1}\\
	S_{(1)(1)}	&=	 S_{(2)(2)}= \frac{(1 - \beta)}{2\beta} S_{(0)(0)},
	\label{eq:stet2}
	\end{align}
 and
	\begin{align}
	{\cal J}_{(0)}	&=	-e^{-(1 - \beta)\phi}[A_{0,z}],\label{eq:currtet1}\\
	{\cal J}_{(1)}	&=	-r^{-1}e^{2\beta\phi}[A_{,z}],
	\label{eq:currtet2}
	\end{align}
respectively. Note that all the quantities in these expressions are evaluated on the surface of the  disk. 

We assume that  there is  no  electric  current  in the  halo,  i.e.  we  take $J_{(a)}^{\pm} 	=0$.  Then,  by using the  procedure for  obtaining  the  metric and  the  electromagnetic  potential developed   in the Secion III  in \cite{G-PGQ2013},  the system under consideration is solvable in terms of various solutions of the Laplace equation. Let us consider the particular case of the Kuzmin solution \cite{kuzmin} such that the metric potential $\phi(r,z)$ is 
	\beq
	\label{solphi}
	\left( 1 + \beta\right) \phi = \ln\left(\frac{\sqrt{r^2 + \left(\mid z \mid + a\right)^2}}{\sqrt{r^2 + \left(\mid z \mid + a\right)^2} - m} \right).
	\eeq
With respect to the LSO, the diagonal components of the three-energy-momentum tensor (evaluated on the surface of the disk), can be interpreted as the energy density and the isotropic pressures of the disk given by the expressions 
	\begin{align}
	\label{energy}
	\rho(r) 	 &= 4\beta ma F(r),\\
	\label{pressure}
	\wp (r) 	& =\frac{(1 -\beta)}{2\beta} \rho(r),
	\end{align}
where the function $F(r)$ is 
	\beq
	\label{fder}
	F(r) = \frac{{(r^2 + a^2)}^{-\frac{2+\beta}{	2 + 2\beta}}}{(1 + \beta ) \left(\sqrt{r^2 + a^2} - m\right)^{\frac{2\beta +1}{1 + \beta}}}. 
	\eeq
Furthermore, the  electromagnetic  field is  determined by  the potential
   \begin{eqnarray}
    A_0(r,z) &=& \frac{k_1}{k} (1 + \beta) \phi	,  \label{eq:electromagnpotA0}\\
    A(r,z) &=-&  \frac{ m |z|\left( (|z| + a) - \sqrt{r^2 + \left(\mid z \mid + a\right)^2} \right) }{k z\sqrt{r^2 + \left(\mid z \mid + a\right)^2}},
    \label{eq:electromagnpotA}\end{eqnarray}
where  $k_1$, $k$, $a$, $m$ and $\beta$ are  arbitrary constants.
Equations (\ref{eq:met0}) - (\ref{eq:electromagnpotA}) completely describe the gravitational and electromagnetic  fields of a thin disk. Notice  that the  disk possesses  an  explicit  magnetic component which  follows from the azimuthal component $A(r,z)$.

From the expressions for the metric and electromagnetic potentials -- equations (\ref{solphi}) and (\ref{eq:electromagnpotA0}), respectively --  the  electric  charge density  on the  surface of  the  disk is given by
 \begin{equation}
 \sigma(r)=\frac{2 a m k_1}{  k (r^2 +  a^2)^{\frac{3 + \beta}{2(\beta +1)}}  \left(\sqrt{r^2 +  a^2}-m\right)^{\frac{2\beta}{\beta +1}}}.
 \end{equation}

Notice that, although the  disk has infinite extension, its mass and charge densities and the azimuthal pressure decay very rapidly  (as $1/r^3$). In every case, the characteristic size can be adjusted through the parameter $a$  of the solution. Moreover, a simple calculation of the curvature invariants reveals that the solution is asymptotically flat. 

\begin{figure}
$$\begin{array}{c}
{\tilde \rho} \\
\epsfig{width=3in,file=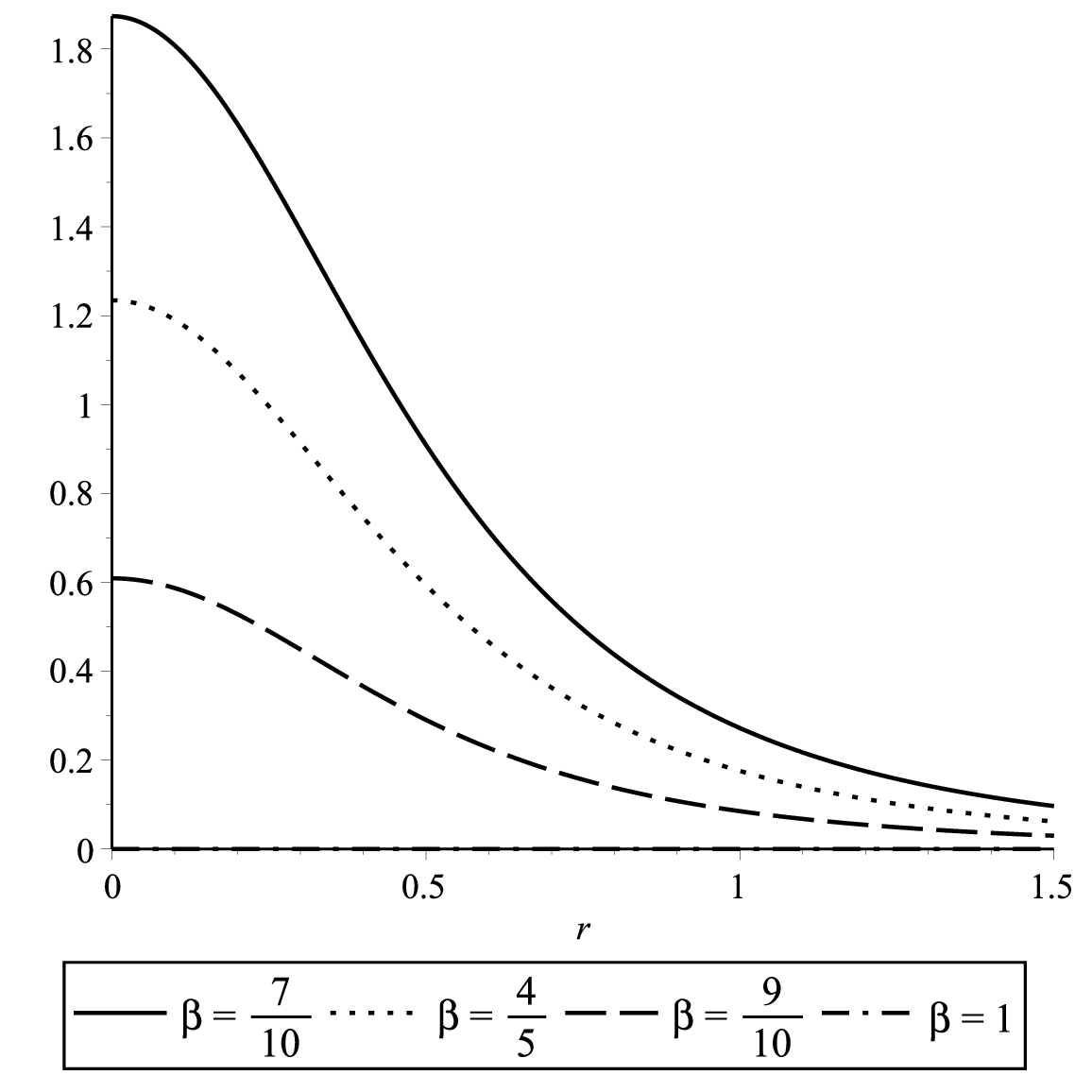}
\label{fig:energy-disk}
\end{array}$$
\caption{\label{fig:figure1} Dimensionless surface energy density ${\tilde \rho}$ as a function of ${\tilde r}$. In each case, we plot ${\tilde
\rho}(\tilde r)$ for ${\tilde m} =0.75$ and different values of the parameter
$\beta $. Here, a tilde indicates division by the solution parameter $a$.  }

\end{figure}

\begin{figure}
$$\begin{array}{c}
{\tilde \sigma} \\
\epsfig{width=3in,file=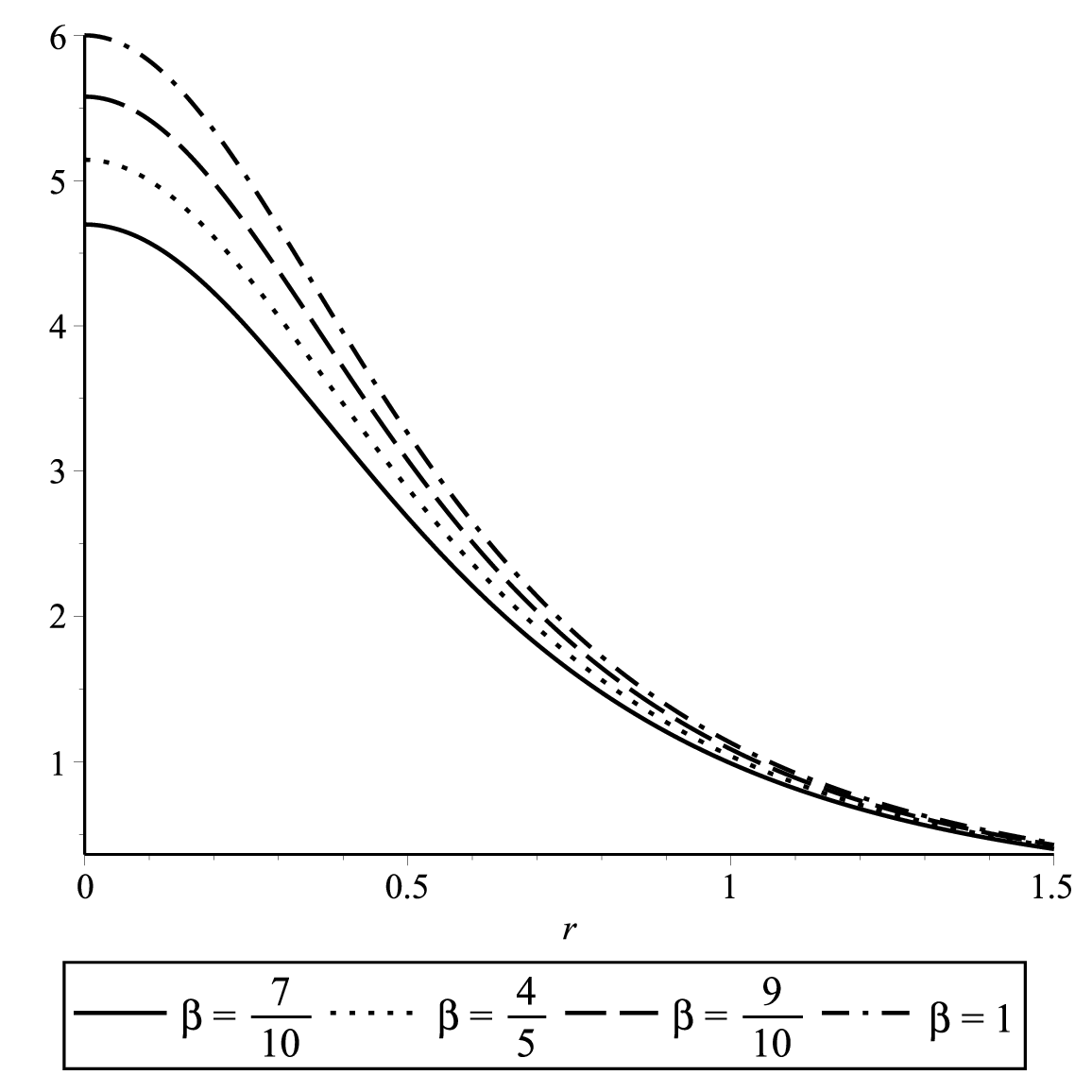}\\
\end{array}$$
\caption{\label{fig:figure2} Dimensionless surface charge density ${\tilde \sigma}$ as a function of ${\tilde r}$. In each case, we plot ${\tilde \sigma}(\tilde r)$  for ${\tilde m} =0.75$, $k_1=k=1$ and different values of the parameter $\beta $.  Here, a tilde indicates division by the solution parameter $a$.}
\label{fig:chargedisk}
\end{figure}

The parameter $\beta$ of the metric produces a non-zero pressure and, from the point of view of the LSO, it has the same value in the radial and angular directions. Note that the particular case of $\beta=1$ corresponds to a preliminary study of this type of solutions found in \cite{G-PGQ2013}. In this work, we are interested in obtaining the thermodynamic properties of a disk described by such type of solutions, for which the case $\beta=1$ is not appropriate. To see this,  we can use the energy conditions on the disk to obtain  the physical range of possible values of the parameter $\beta$, these values are shown in table \ref{table1}.

		\begin{table}
		\begin{tabular}{|l|c|c|}
		\hline
		WEC			&  $\rho \geq 0$, $\rho + {\wp} > 0$		  						& $\beta \in  (-\infty, -1] \cup [0, \infty)$\\
		\hline
		NEC			&  $\rho + {\wp} \geq 0$,											& $1\geq 0$ (no information)\\
		\hline
		SEC			&  $\rho + 2{\wp} \geq 0$, $\rho + {\wp} \geq 0$					& $\beta \in [-1, \infty)$\\
		\hline
		DEC 		&  $\rho  \geq 0$, $\rho   \geq  |{\wp}|$							& $\beta \in (-\infty, -1]\cup [{1}/{3}, \infty)$\\
		\hline
		\end{tabular}
		\caption{The range of $\beta$ for which the energy conditions hold. }
		\label{table1}
		\end{table}

Moreover, we can read off the value of the adiabatic speed of sound for the LSO from equation \eqref{pressure}. It follows that, in order to satisfy simultaneously the energy conditions and the causality requirement for the speed of sound, we must require that $\beta\in (1/3,\infty)$.

The main point we want to address here arises from the fact that we can use the variational techniques for relativistic thermodynamics \cite{carter,heatpaper} to obtain the thermal properties of the material content of the disk. The central role in such a formalism is played by the so-called `master function' $\Lambda$, the Lagrangian density of the  matter content in the Einstein-Hilbert action. Let us assume that the disk consists of a multicomponent fluid described by its particle number and entropy density currents. Since the configuration at hand is (conforma)static, these currents cannot depend on time and must be aligned with the time-like Killing vector field of the metric. Thus, with respect to the LSO, the multifluid components are given by
	\beq
	n^a = n(r) \delta^{\ a}_{0} \quad \text{and}  \quad s^a = s(r) \delta^{\ a}_{0},
	\eeq
respectively. The leading role in the multifluid formalism is not played by the currents, but by their corresponding conjugate momenta 
	\beq
	\mu_a = \frac{\partial \Lambda}{\partial n^a} \quad  \text{and}  \quad \theta_a = \frac{\partial \Lambda}{\partial s^a}.
	\eeq
Thus, the currents and momenta are completely specified by their time-like components which we assume to be functions of $r$ alone, namely $n(r)$ and $s(r)$.

Substituting the variational definition of the energy-momentum tensor (c.f. equation (2.13) in \cite{heatpaper}), namely
	\beq
	T_a^{\ b} = \mu_a n^b + \theta_a s^{b} + \Psi \delta_a^{\ b},
	\eeq
 where $\Psi = \Lambda - \mu_a n^a - \theta_a s^a$ is the multifluid generalised pressure, as the source of the Einstein field equations restricted to the disk surface and using the solution \eqref{solphi}, it follows that the master function is simply 
	\beq
	\Lambda = -\rho(r),
	\eeq
in agreement with the definition of local thermodynamic equilibrium \cite{heatpaper} which is a consequence of the time symmetry of the present solution. We also obtain a single differential equation stemming from the identification of  the generalised pressure $\Psi$ with the pressure measured by the LSO $\wp$ [c.f. equation \eqref{pressure}]. Thus, the single equation relating the energy with the particle number ande entropy densities is
	\beq
	\label{efe.03}
	\left(\frac{\d}{\d r}\ln\rho\right) \left(\frac{\d}{\d r} n s\right) = \left(\frac{1 + \beta}{2\beta} \right)\frac{\d n}{\d r}  \frac{\d s}{\d r} .
	\eeq	
This equation admits solutions of the form
	\beq
	\label{sol1}
	n  = A \rho^{\kappa_\n} \quad \text{and} \quad 	s  = B \rho^{\kappa_\s},
	\eeq
where $A$, $B$, $\kappa_\n$ and $\kappa_\s$ are constants, and $\kappa_\n$ and $\kappa_\s$ must satisfy the relation
	\beq
	\label{efep}
	\frac{\kappa_\n + \kappa_\s}{\kappa_\n \kappa_\s} =  \frac{1 + \beta}{2\beta},
	\eeq
constraining the type of matter compatible with the solution. In particular, the case where $\kappa_\n = 1$ and $\kappa_\s = 3/4$ corresponds to electromagnetically charged  dust and the solution is completely determined by \eqref{sol1} with $\beta=3/11$.  To verify this, let us compute the adiabatic speed of sound of each component (c.f. equation (29) in \cite{twostream})
	\beq
	\label{ssn}
	c_i^2  = \frac{1 - \kappa_i}{\kappa_i}, \quad \text{with} \quad i=\n,\s.
	\eeq
Thus, for the constants we have chosen, we have $c_\n^2 = 0$ and $c_\s^2 = 1/3$, in agreement with our physical interpretation. Note however, that the sole value of $\beta$ does not uniquely determine the type of matter for the disk. Moreover, from equations \eqref{ssn}, it is straightforward to obtain  the range of $\beta$ for which the multi-fluid interpretation is causal.   That is, $1/2 \leq \kappa_i \leq 1$, and therefore $ 1/5 \leq\beta \leq 1/3$. This result implies that the single fluid interpretation of the system is incompatible with the multi-fluid thermodynamic description.  We now show that the latter has a richer physical content in the sense that the thermodynamics associated with the multi-fluid model is consistent with the matter content assumed for the disk.

Integrating  equations \eqref{ssn} and substituting $\Lambda = -\rho$ we obtain the material fundamental thermodynamic relation 
	\beq
	\label{frel}
	\rho(n,s) = \mu_0 n^{1/\kappa_\n} + \theta_0 s^{1/\kappa_\s},
	\eeq
where $\mu_0$ and $\theta_0$ are integration constants. Now, projecting each conjugate momenta into the LSO frame we get the chemical potential and temperature, respectively \cite{heatpaper}. Thus, using the fundamental relation \eqref{frel} and the solutions \eqref{sol1} we obtain
	\begin{align}
	\label{mom1}
	\mu 	&= \frac{\mu_0}{\kappa_\n} A^{\frac{1- \kappa_\n}{\kappa_\n}} \rho^{1- \kappa_\n},\\
	\label{mom2}
	T 		&= \frac{\theta_0}{\kappa_\s} B^{\frac{1- \kappa_\s}{\kappa_\s}} \rho^{1- \kappa_\s}.
	\end{align}
In the case of  $\kappa_\n=1$ and $\kappa_\s= 3/4$ we observe that the chemical potential of the dust is constant across the disk and the temperature is proportional to $\rho^{1/4}$. 

\begin{figure}
\includegraphics[width=0.49\columnwidth]{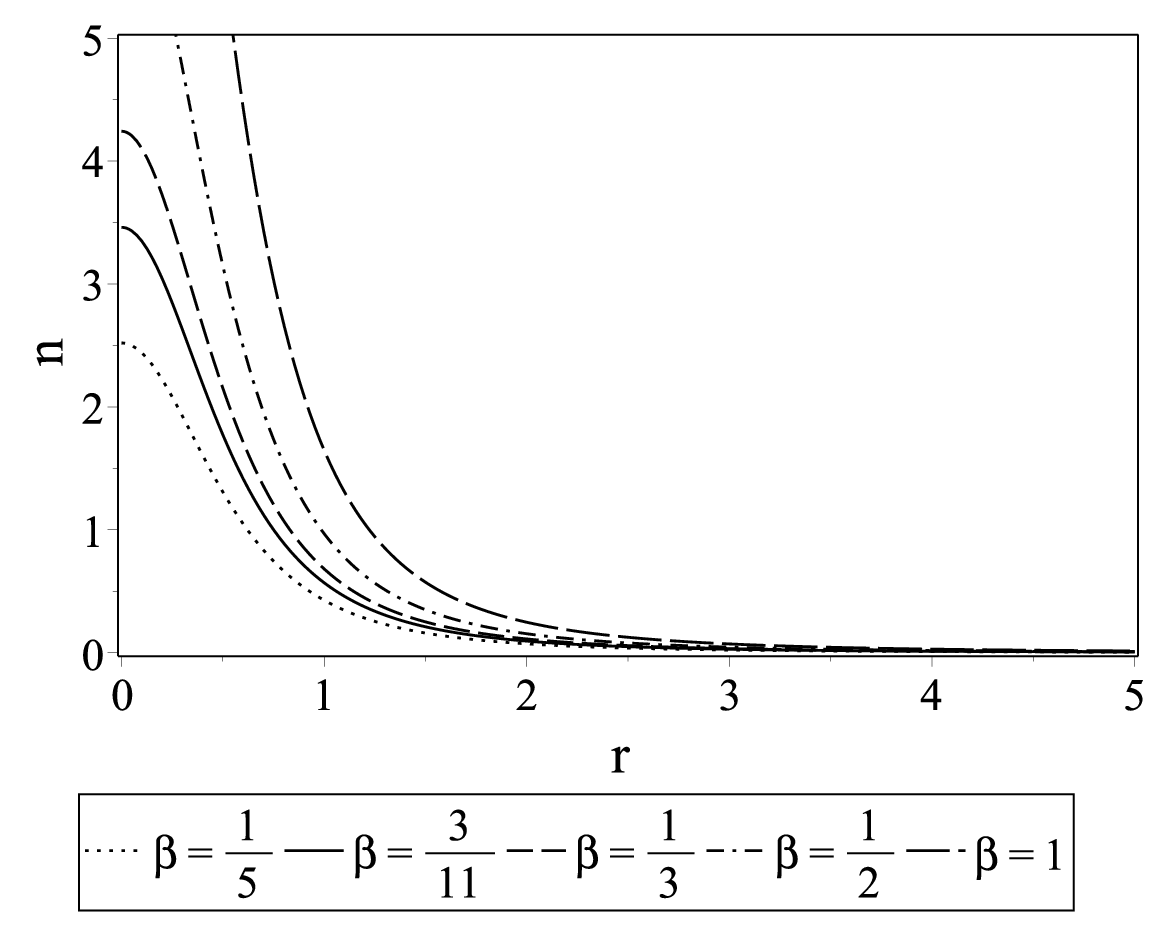}  \includegraphics[width=0.49\columnwidth]{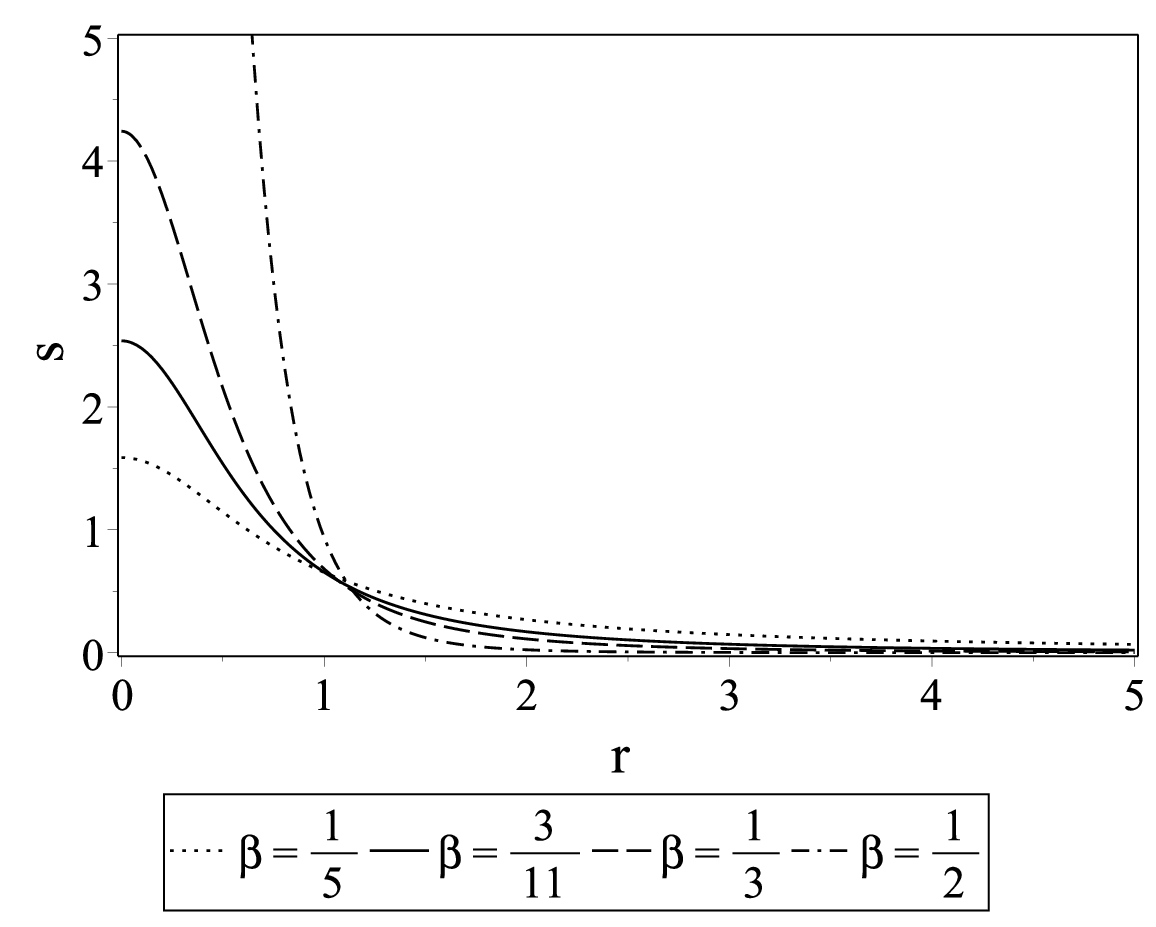}\\
\includegraphics[width=0.49\columnwidth]{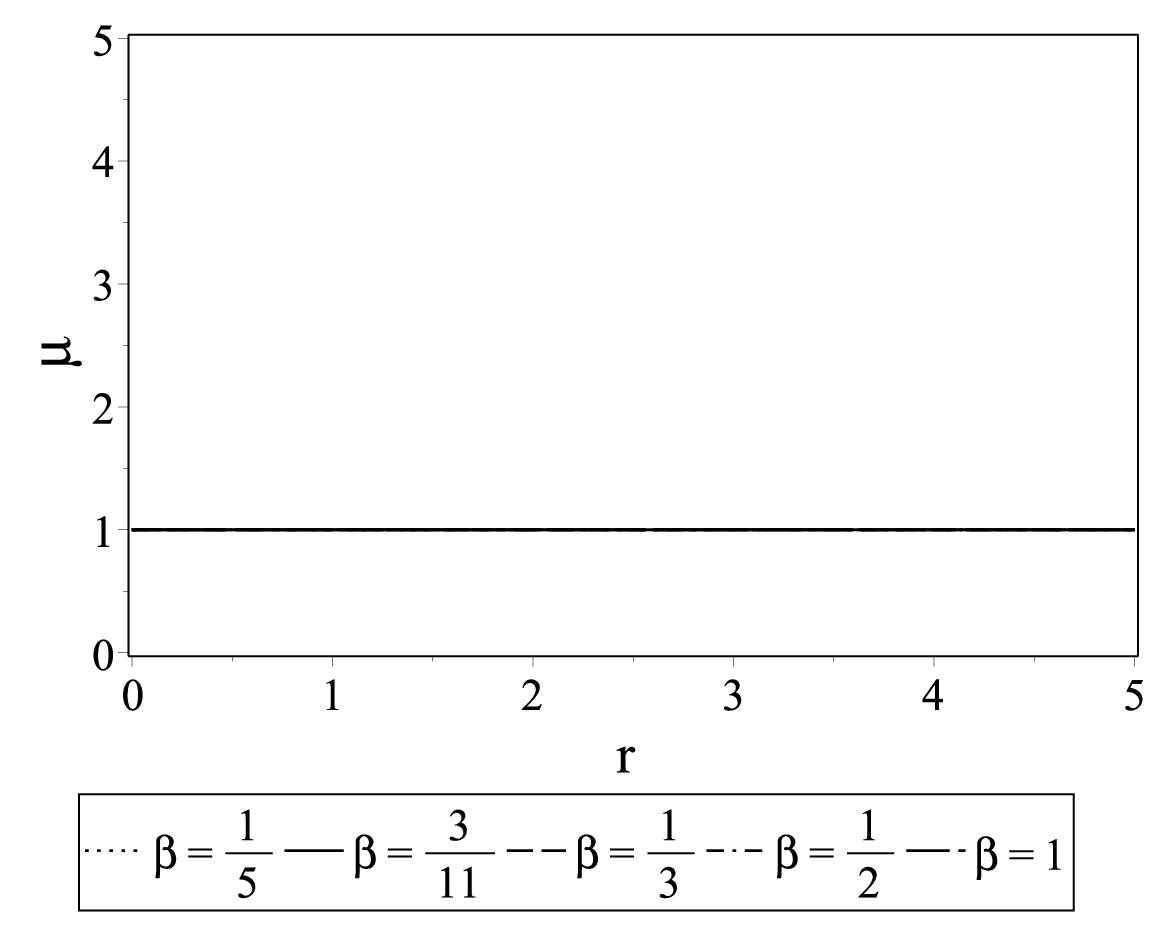}  \includegraphics[width=0.49\columnwidth]{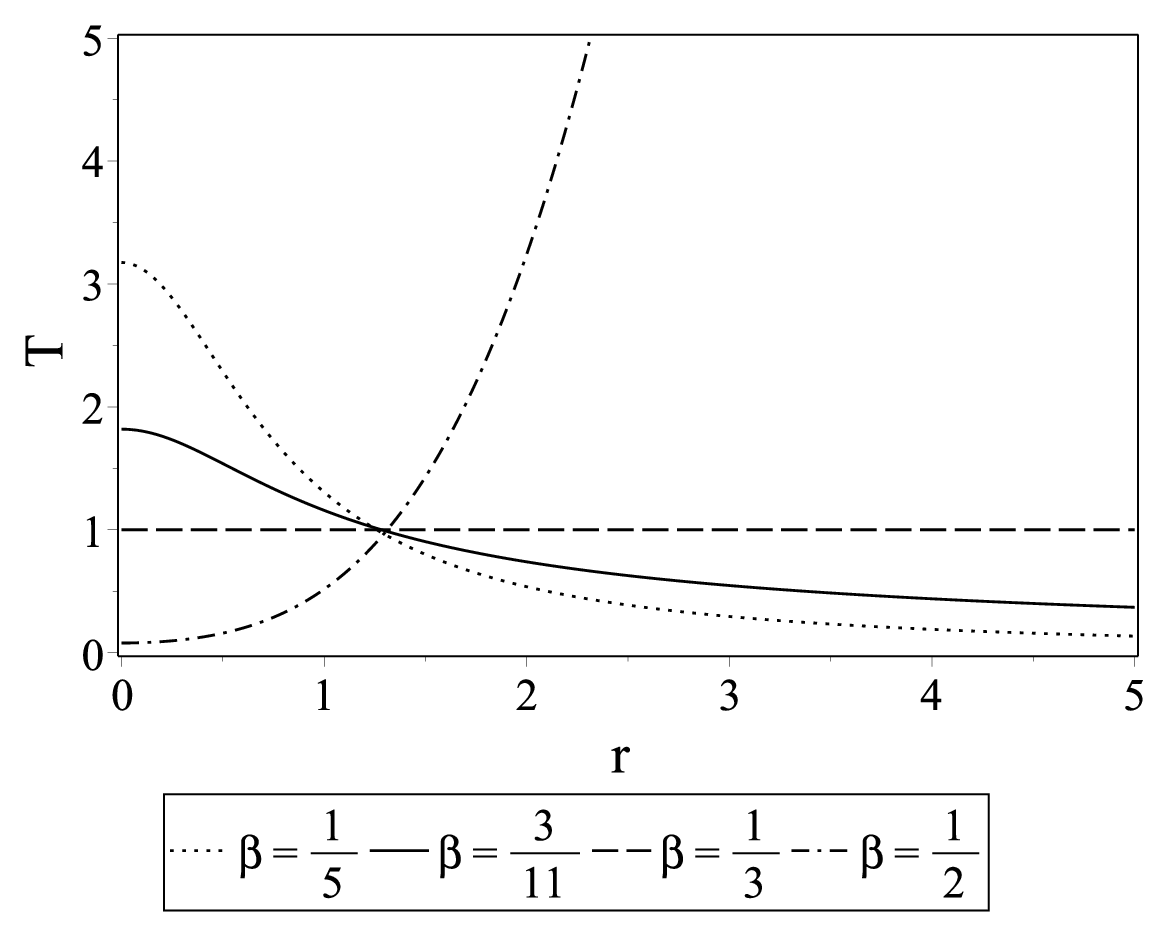}
\caption{Thermodynamic quantities on the disk for different values of $\beta$. In the top part we show the particle number and entropy densities whilst in the bottom their corresponding conjugate quantities, chemical potential and temperature. The plots in the left are well behaved for any value of $\beta \in [1/5,1]$, the chemical potential is constant. The plots on the right are not defined for the case $\beta =1$. Moreover there are divergences for values of $\beta >1/3$. In all cases, the solid line represents the electromagnetic case we study. Here, we have used the values $a = 1$ and $m = 0.75$ [c.f. equation \eqref{solphi}]. }
\label{fig1}
\end{figure}

The solutions \eqref{sol1} together with their thermodynamic conjugate quantities \eqref{mom1} and \eqref{mom2}, satisfy Euler's identity $\rho + \Psi = \mu n + T s$, for any value of $\beta$. In figure \ref{fig1} we show the various thermodynamic quantities [equations \eqref{sol1} and \eqref{mom1} -\eqref{mom2}] corresponding to different values of $\beta$.

This exercise shows that the scheme presented here is consistent with the physical input we considered. Furthermore, this solution is suitable for a vast family of master functions (fundamental relations) for the material content of the disk provided the relation between the constants, equation \eqref{efep}, is satisfied. Thus for multicomponent models for which one component is proportional to the energy density of the disk ($\kappa_\n=1$), the pressure must be generated by the other component. In the multi-fluid thermodynamic interpretation, this component is typically identified with the entropy of the disk \cite{heatpaper}  and is controlled by the parameter $\kappa_\s$. Thus, fixing the particular solution compatible with the matter content.     

In sum, we have presented an exact solution for modelling relativistic thin disks. The relevance of the solution is essentially two-fold. On the one hand, it has a remarkably simple mathematical form. The solution  \eqref{solphi} determines the behaviour of the material content through the function $F(r)$, equation \eqref{fder}. On the other hand, the multi-fluid interpretation of the solution has allowed us for the first time to give a complete thermodynamic description of the system in terms of two parameters which determine the matter content of for the disk. It remains to give a complete thermodynamic treatment for the halo.  This work serves as a `proof of principle' that gives a solid footing for a fuller study of relativistic disks, in particular, for a later study of the more realistic stationary solution.

A.C.G-P. is thankful to Departamento de Gravitaci\'on y Teor\'ia  de Campos (ICN-UNAM) for its hospitality and kind support though the course of his post-doctoral fellowship. He also wants to thank  COLCIENCIAS, Colombia, and TWAS-CONACYT for financial support. CSLM acknowledges  support from a UNAM-DGAPA postdoctoral grant. HQ receives partial support from  CONACYT, Grant No. 166391 and DGAPA-UNAM. The authors would like to thank Francisco Nettel for useful comments and discussions.



\appendix
\section{} 

 For  the  metric (\ref{eq:met0}) the  non-zero components  of  the energy-momentum  tensor  of  the halo  as observed by a LSO are  given by
	\begin{align}
	{M_{(0)(0)}^{\pm}} 	&= 	-e^{2\beta\phi}\{-2\beta\nabla^2\phi + \beta^2 \nabla\phi\cdot\nabla\phi	+\frac{1}{2}e^{-2\phi}\nabla A_0\cdot\nabla A_0 + \frac{1}{2}r^{-2}e^{2\beta\phi}\nabla A\cdot\nabla A	\},\label{eq:emth2mP}\\
	{M_{(0)(1)}^{\pm}} 	&= 	-r^{-1}e^{-(1 -3\beta)\phi} \nabla A_0 \cdot \nabla	A,\\
	{M_{(1)(1)}^{\pm}} 	&= 	e^{2\beta\phi}\{(1 - \beta )\nabla^2{\phi} - (1 -\beta )\frac{1}{r}\phi_{,r} + \nabla{\phi}\cdot\nabla{\phi} -\frac{1}{2}e^{-2\phi}\nabla A_0\cdot\nabla A_0 -	\frac{1}{2}r^{-2}e^{2\beta\phi}\nabla A\cdot\nabla A\},\\
	{M_{(2)(2)}^{\pm}} 	&= 	e^{2\beta\phi}\{ (1 - \beta )\nabla^2{\phi} - (1 -\beta )\phi_{,rr} + \phi_{,z}^2 + (\beta^2 -2\beta)\phi_{,r}^2 + \frac{1}{2}e^{-2\phi}(A_{0,r}^2 -A_{0,z}^2) -	\frac{1}{2}r^{-2}e^{2\beta\phi}(A_{,r}^2- A_{,z}^2)\},\\
	{M_{(2)(3)}^{\pm}} 	&= 	e^{2\beta\phi}\{(\beta -1)\phi_{,rz} + (\beta^2 - 2\beta - 1)\phi_{,r}\phi_{,z} + e^{-2\phi}A_{0,r}A_{0,z} - r^{-2}e^{2\beta\phi}A_{,r}A_{,z}  \},\label{eq:nondiagm23}\\
	{M_{(3)(3)}^{\pm}} 	&=	e^{2\beta\phi}\{ (1 - \beta )\nabla^2{\phi} - (1 -\beta )\phi_{,zz} + \phi_{,r}^2 + (\beta^2 -2\beta)\phi_{,z}^2 - \frac{1}{2}e^{-2\phi}(A_{0,r}^2 -A_{0,z}^2) + \frac{1}{2}r^{-2}e^{2\beta\phi}(A_{,r}^2- A_{,z}^2)\}.\label{eq:emth2mU}
	\end{align}

\end{document}